# Directed Threshold Multi – Signature Scheme without SDC

**Sunder Lal** [*] **and Manoj Kumar** [**]

*\* Dept of Mathematics, IBS Khandari.Dr. B.R.A.University Agra.*
Sunder_lal2@rediffmail.com.in.

*\*\* Dept of Mathematics, HCST, Farah – Mathura, [U. P.] – 281122.*
yamu_balyan@yahoo.co.in

**Abstract.** In this paper, we propose a **Directed threshold multisignature scheme without SDC.** This signature scheme is applicable when the message is sensitive to the signature receiver; and the signatures are generated by the cooperation of a number of people from a given group of senders. In this scheme, any malicious set of signers cannot impersonate any other set of signers to forge the signatures. In case of forgery, it is possible to trace the signing set.

## 1. Introduction

Digital signature is a cryptographic tool to authenticate electronic communications. Digital signature scheme allows a user with a public key and a corresponding private key to sign a document in such a way that anyone can verify the signature on the document (using her/his public key), but no one can forge the signature on any other document. This self-authentication is required for some applications of digital signatures such as certification by some authority.

In most situations, the signer is generally a single person. However, in some cases the message is sent by one organization and requires the approval or consent of several people. In these cases, the signature generation is done by more than one consenting person. A common example of this policy is a large bank transaction, by one organization, which requires the signature of more than one partner. Such a policy could be implemented by having a separate digital signature for every required signer, but this solution increases the effort to verify the message linearly with the number of signer. To solve this problems, Multisignature schemes [10,11,13,17,18] and threshold signature schemes [5,6,7,9,12,20] are used where more than one signers share the responsibility of signing messages

Threshold signatures are closely related to the concept of threshold cryptography, first introduced by Desmedt [5, 6, 7]. In 1991, Desmedt and Frankel [6] proposed the first (t, n) threshold digital signature scheme based on the RSA system. In ($t, n$) threshold signature scheme, any subgroup of $t$ or more shareholders of the designated group can generate a valid group signature in such a way that the verifier can check the validity of the signature without identifying the identities of the signers. In threshold schemes, when any $t$ or more shareholders act in collusion, they can impersonate any other set of shareholders to forge the signatures. In this case, the malicious set of signers does not have any responsibility for the signatures and it is impossible to trace the signers. Unfortunately, with threshold schemes proposed so far, this problem cannot be solved.

In multisignature schemes, the signers of a multisignature are identified in the beginning and the validity of the multisignature has to be verified with the help of identities of the signers. For multisignature, it is indeed unnecessary to put a threshold value to restrict the number of signers. Consider the situation, where a group of anonymous members would have to generate a multisignature. The members of this group use pseudonyms as their identities in the public directory. What concerns the verifier most is that at least t members sign a message and they indeed come from that group. Nevertheless, the verifier has no way to verify whether a user is in fact a member of that group because of the anonymity of the membership. In this case, the multisignature schemes cannot solve this problems, however, the threshold signature schemes do.

On the other hands, there are so many situations, when the signed message is sensitive to the signature receiver. Signatures on medical records, tax information and most personal/business transactions are such situations. Signatures used in such situations are called directed signatures [1, 2, 3, 14, 15, 19, 23, 24]. In directed signature scheme, the signature receiver has full control over the signature verification process and can prove the validity of the signature to any third party, whenever necessary. Nobody can check the validity of signature without his cooperation.

Many threshold signature schemes require a **trusted SDC** to generate the group secret keys and secret shares of group members, which have a single point of vulnerability. The existence of such a center is not a reasonable assumption; there are two potential problems.

- First, for many applications, there is no one person or devices which cab be completely trusted by all members of the group.
- Second, the use of a key center creates a single point failure. Any security lapse at the key center can reveal the private key.

To avoid these problems, in 1992, Harn introduced a scheme based on a modified ElGamal signature scheme, which does not require a trusted SDC [10]. Each member works as a SDC, generates and distributes the secret shares for each user. In this paper, we proposed a *Directed - Threshold Multi - Signature Scheme without SDC.*

## 2. Directed - Threshold Multi - Signature Scheme without SDC

In this scheme, each shareholder works as a SDC to generate his secret key and distribute the corresponding secret shares to other shareholders. In advance, all the shareholders are agree for the public parameters. We assume that there is a **designated combiner *DC*** who takes the responsibility to collect and verify each partial signature and then produce a group signature. **Nevertheless, there is no partial secret information of the other users associated with the *DC*.**

### 2.1. Group Public Key and Secret Shares Generation Phase

(a). All the members of the organization $S$ agree to select the group public parameters $p, q, g, h$ and a common random secret $K$.

(b). Each member $i$ in $G_S$ computes $W = g^{-K} \mod p$.

(c). Each member $i$ in $G_S$ randomly selects a $(t-1)^{th}$ degree polynomial $f_i(x)$ secretly and an integer $u_{S_i} \in Z_q$ publicly.

(d). Each member $i$ in $G_S$ computes partial group public key $y_i = g^{f_i(0)} \mod p$.

(e). The group public key $y_S$ is given by $y_S = \prod_{i \in G_S} y_i \mod p$.

(f). Each member $i$ in $G_S$ works as SDC. He selects a random secret $h_{ij} \in Z_q$ and computes the secret value $l_{ij}$ and public value $m_{ij}, n_{ij}$ for $j \neq i$, as,

$$l_{ij} = [h_{ij} + f_i(u_{S_i})] \mod q,$$

$$m_{ij} = g^{l_{ij}} \mod p \text{ and } n_{ij} = g^{h_{ij}} \mod p.$$

(g). Each member $i$ in $G_S$ computes a public value $v_{S_i}$ for each member $j \neq i$, of the group $G_S$, as, $v_{S_{ij}} = l_{ij} \cdot y_{S_j}^{K} \mod p$.

Here, $y_{S_j}$ is the public key associated with each user $j$ in the group $G_S$.

(h). Each member $i$ in $G_S$ sends $v_{S_{ij}}$ to each member $j \neq i$, through a public channel.

### 2.2. Partial Signature Generation by any t Members and Verification

If any $t$ members of the organization out of $n$ members agree to sign a message $m$ for a person R. R possesses a pair ($x_R, y_R$). Then the signature generation has the following steps.

(a). Each member $i, i \in H_S$, randomly selects $K_{i_1}$ and $K_{i_2} \in Z_q$ and computes

$$u_i = g^{-K_{i_2}} \mod p, \quad v_i = g^{K_{i_1}} \mod p \text{ and } w_i = g^{K_{i_1}} y_R^{K_{i_2}} \mod p.$$

(b). Each member makes $u_i, w_i$ publicly and $v_i$ secretly available to each member of $H_S$. Once all $u_i, v_i$ and $w_i$ are available, each member $i, i \in H_S$ computes the product $U_S, V_S, W_S$ and a hash value $R_S$, as,

$$U_S = \prod_{i \in H_S} u_i \bmod q, \quad V_S = \prod_{i \in H_S} v_i \bmod q,$$

$$W_S = \prod_{i \in H_S} w_i \bmod q, \text{ and } R_S = h(V_S, m) \bmod q.$$

(c). Each member $i, i \in H_S$ recovers his/her secret share $l_{ji}$, as,

$$l_{ji} = v_{S_{ji}} W^{x_{S_i}} \bmod p. \quad (j \neq i.)$$

(d). Each member $i, i \in H_S$ computes a value $C_i = \prod_{k \in H_S, k \neq i} \frac{(0 - u_{S_k})}{(u_{S_i} - u_{S_k})} \bmod q.$

(e). Each member of $H_S$ computes his/her modified shadow $MS_{S_i} = \sum_{j \in G_S, j \notin H_S} l_{ji} \cdot C_i \bmod q.$

(f). Each member $i, i \in H_S$ uses his/her modified shadow, $MS_{S_i}$ and a value $s_i$, as

$$s_i = [K_{i_1} + (f_i(0) + MS_{S_i}) \cdot R_S] \bmod q.$$

(g). Each member $i, i \in H_S$ sends his partial signature to the designated combiner $DC$. $DC$ verify the partial signature ($s_i, v_i, C_i, R_S$) by the relation,

$$g^{s_i} \stackrel{?}{\equiv} v_i \cdot \left( y_i \prod_{j \in G_S, j \notin H_S} m_{ji}^{C_i} \right)^{R_S} \bmod p.$$

If the above equation holds, then the partial signature($s_i, v_i, C_i, R$) for shareholder $i$ is valid.

**2.3. Group Signature Generation.**

(a). $DC$ can computes the group signature $S_S = \sum_{i=1}^{t} s_i \bmod q$ by combining all the partial signature.

(b). $DC$ sends $\{S_S, U_S, W_S, m\}$ as signature of the group $S$ for the message $m$ to $R$.

*2.4. Signature Verification by R*

To verify the validity of the group signature $\{S_S, U_S, W_S, m\}$ **the verifier $R$ needs is/her secret key** $x_R$. This sub-section consists of the following steps.

(a). The verifier $R$ computes a verification value $E = \prod_{i \in H_S} \left( \left( \prod_{j \in G_S, j \notin H_S} n_{ji} \right)^{C_i} \right) \bmod p.$

(b). Only the verifier $R$ can recovers the values $R_R$ and $R_S$, as

$$R_R = W_S \cdot U_S^{x_R} \bmod p \text{ and } R_S = h(V_S, m).$$

(c). The verifier $R$ uses the congruence $g^{S_S} \overset{?}{\equiv} R_R \cdot (E \cdot y_S)^{R_S} \bmod p$ to check the validity the signature. If this congruence holds, then the group signature $\{S_S, U_S, W_S, m\}$ is valid signature of the organization $S$ on the message $m$.

## 2.5. Proof of Validity by R to any third Party C

(a). $R$ computes $\mu = U_S^{x_R} \bmod p$ and $R_R = \mu \cdot W_S \bmod p$.

(b). $R$ sends $\{R_R, E, S_S, U_S, m, \mu\}$ to C.

(c). C recovers $R_S = h(R_R, m)$ and uses the following congruence to check the validity of the signature

$$g^{S_S} \overset{?}{\equiv} R_R \cdot (E \cdot y_S)^{R_S} \bmod p.$$

If this does not hold C stops the process; otherwise goes to the next step.

(d) In a zero knowledge fashion $R$ proves to C that $\log_{U_S} \mu = \log_g y_R$ as follows:-

- C chooses random $u, v \in Z_p$ computes $w = (U_S)^u \cdot g^v \bmod p$ and sends $w$ to R.
- $R$ chooses random $\alpha \in Z_p$ computes $\beta = w \cdot g^\alpha \bmod p$, $\gamma = \beta^{x_R} \bmod p$ and sends $\beta, \gamma$ to C.
- C sends $u, v$ to R, by which $R$ can verify that $w = (U_S)^u \cdot g^v \bmod p$.
- $R$ sends $\alpha$ to C, by which she can verify that

$$\beta = (U_S)^u \cdot g^{v+\alpha} \bmod p, \quad \text{and} \quad \gamma = (\mu)^u \, y_R^{v+\alpha} \bmod p.$$

## 3. Security Discussions

In this sub-section, we shall discuss the security aspects of proposed scheme. Here we shall discuss several possible attacks and show that, none of these can successfully break the system.

(a). Is it possible to retrieve the partial secret keys $f_i(0)$, $i \in G_S$ ?

This is as difficult as solving discrete logarithm problem. No one can get the partial group public key $y_i$, since $f_i$ is the randomly and secretly selected polynomial by the member $i$. On the other hand, by using the public keys $y_S$ no one also get the partial secret keys $f_i(0)$ because

$$y_i = g^{f_i(0)} \bmod p \text{ and } y_S = \prod_{i \in G_S} y_i \bmod p.$$

(b). Is it possible to retrieve the secret share $l_{ij}$ from the equation

$$l_{ij} = [h_{ij} + f_i(u_{S_i})] \bmod q \ ?$$

No, because $f_i$ is the randomly and secretly selected polynomial and $h_{ij}$ is also a randomly and secretly selected integer by the member $i$.

(c). Is it possible to retrieve the secret share $l_{ij}$ from the equation

$$m_{ij} = g^{l_{ij}} \mod p \ ?$$

No, because this is as difficult as solving discrete logarithm problem.

(d). Is it possible to retrieve the secret share $l_{ij}$ from the equation

$$v_{S_{ij}} = l_{ij} \cdot y_{S_j}{}^K \mod p?$$

No because $K$ is a randomly and secretly selected common integer.

(e). Is it possible to retrieve the secret shares $l_{ij}$ from the equation

$$l_{ji} = v_{S_{ji}} W^{x_{S_i}} \mod p?$$

Only the user $i$ can recovers his secret shares $l_{ij}$ because $x_{S_i}$ is secret key of the user $i$.

(f). Can one retrieve the modified shadow $MS_{S_i}$ from the equation

$$MS_{S_i} = \sum_{j \in G_S, j \notin H_S} l_{ji} \cdot C_i \mod q \ ?$$

It is impossible to collect the modified shadow $MS_{S_i}$ from the equation because all $l_{ij}$ are secret information shared by the users.

(g). Is it possible that the designated combiner $DC$ retrieve the any partial information from the equation

$$S_S = \sum_{i \in H_S} s_i \mod q \ ?$$

Obviously, this is computationally infeasible for $DC$.

(h). Is it possible to that any one can impersonate a user $i \in H$ ?

A forger may try to impersonate a user $i \in H_S$, by randomly selecting two integers $K_{i_1}$ and $K_{i_2} \in Z_q$ and broadcasting $u_i$, $v_i$ and $w_i$. But without knowing the secret shares $l_{ij}$ and $R_S$, it is difficult to generate a valid partial signature $s_i$ to satisfy the verification equations,

$$g^{s_i} \stackrel{?}{\equiv} v_i \cdot \left( y_i \prod_{j \in G_S, J \notin H_S} m_{ji}{}^{C_i} \right)^{R_S} \mod p.$$

(g). Is it possible to that any one can forge a signature $\{S_S, U_S, W_S, m\}$ by the following equation

$$g^{S_S} \stackrel{?}{\equiv} R_R \cdot (E \cdot y_S)^{R_S} \mod p \ ?$$

A forger may randomly selects an integer $R_R$ and then computes the hash value $R_S$ such that $R_S = h(R_R, m) \mod q$, obviously to compute the integer $S_S$ is equivalent to solving the discrete logarithm problem. On the other hand, the forger can randomly select $R_S$ and $S_S$ first, then try to determine a value $R_R$ that satisfies the signature verification equation. However, according to the property of the hash function $h$, it is quite impossible. Thus, this attack will not be successful.

## 4. Illustration

For illustration, Suppose $|G_S| = 7$, $|H_S| = 4$, $p = 47$, $q = 23$, $g = 3$, $K = 11$ and $W = 12$.

### Group public Key and Secret Shares Generation Phase

(a). All the users compute their secret and public values, which are given by the following table.

| VALUE USER | Secret $f_i(x)$ | Secret $f_i(0)$ | Public $u$ | Public $y_i$ | Secret $x_{S_i}$ | Public $y_{S_i}$ |
|---|---|---|---|---|---|---|
| User-$S_1$ | $7 + 12 x^3$ | 7 | 9 | 25 | 9 | 37 |
| User-$S_2$ | $9 + 11 x^3$ | 9 | 13 | 37 | 11 | 4 |
| User-$S_3$ | $14 + 8 x^3$ | 14 | 15 | 14 | 13 | 36 |
| User-$S_4$ | $17 + 3 x^3$ | 17 | 11 | 2 | 19 | 18 |
| User-$S_5$ | $13 + 7 x^3$ | 13 | 18 | 36 | 5 | 8 |
| User-$S_6$ | $18 + 15x^3$ | 18 | 19 | 6 | 10 | 17 |
| User-$S_7$ | $21 + 15 x^3$ | 21 | 21 | 21 | 14 | 14 |

(b). The group public key $y_S$ is given by $y_S = 25$.

(c). User-$S_1$ makes the following table.

| VALUE USER | secret $h_{1j}$ | secret $l_{1j}$ | public $m_{1j}$ | public $n_{1j}$ | public $v_{1j}$ |
|---|---|---|---|---|---|
| User – $S_2$ | 14 | 4 | 34 | 14 | 2 |
| User – $S_3$ | 9 | 13 | 36 | 37 | 10 |
| User – $S_4$ | 11 | 5 | 8 | 4 | 45 |
| User – $S_5$ | 15 | 17 | 2 | 42 | 18 |
| User – $S_6$ | 17 | 15 | 42 | 2 | 43 |
| User – $S_7$ | 13 | 16 | 32 | 36 | 42 |

(d). User- $S_2$ makes the following table.

| VALUE\USER | Secret $h_{2j}$ | Secret $l_{2j}$ | Public $m_{2j}$ | Public $n_{2j}$ | Public $v_{2j}$ |
|---|---|---|---|---|---|
| User – $S_1$ | 13 | 14 | 14 | 36 | 21 |
| User – $S_3$ | 14 | 3 | 27 | 14 | 24 |
| User – $S_4$ | 9 | 8 | 28 | 17 | 25 |
| User – $S_5$ | 7 | 21 | 21 | 25 | 25 |
| User – $S_6$ | 16 | 11 | 4 | 8 | 19 |
| User – $S_7$ | 3 | 16 | 32 | 27 | 42 |

(e). User- $S_3$ and makes the following table.

| VALUE\USER | Secret $h_{3j}$ | Secret $l_{3j}$ | Public $m_{3j}$ | Public $n_{3j}$ | Public $v_{3j}$ |
|---|---|---|---|---|---|
| User – $S_1$ | 7 | 11 | 4 | 25 | 40 |
| User – $S_2$ | 10 | 5 | 8 | 17 | 26 |
| User – $S_4$ | 11 | 1 | 3 | 4 | 9 |
| User – $S_5$ | 13 | 16 | 32 | 36 | 28 |
| User – $S_6$ | 19 | 4 | 34 | 18 | 24 |
| User – $S_7$ | 21 | 17 | 2 | 21 | 27 |

(f). User- $S_4$ makes the following table.

| VALUE\USER | Secret $h_{4j}$ | Secret $l_{4j}$ | Public $m_{4j}$ | Public $n_{4j}$ | Public $v_{4j}$ |
|---|---|---|---|---|---|
| User – $S_1$ | 19 | 15 | 14 | 18 | 46 |
| User – $S_2$ | 13 | 20 | 7 | 36 | 10 |
| User – $S_3$ | 11 | 10 | 17 | 4 | 33 |
| User – $S_5$ | 6 | 16 | 32 | 24 | 28 |
| User – $S_6$ | 8 | 17 | 2 | 28 | 8 |
| User – $S_7$ | 18 | 11 | 4 | 6 | 23 |

(g). User- $S_5$ makes the following table.

| VALUE \ USER | Secret $h_{5j}$ | Secret $l_{5j}$ | Public $m_{5j}$ | Public $n_{5j}$ | Public $v_{5j}$ |
|---|---|---|---|---|---|
| User – $S_1$ | 6 | 16 | 32 | 24 | 24 |
| User – $S_2$ | 17 | 22 | 16 | 2 | 11 |
| User – $S_3$ | 18 | 15 | 42 | 6 | 26 |
| User – $S_4$ | 13 | 5 | 8 | 36 | 45 |
| User – $S_6$ | 19 | 21 | 21 | 18 | 32 |
| User – $S_7$ | 8 | 11 | 4 | 28 | 23 |

(h). User- $S_6$ makes the following table

| VALUE \ USER | Secret $h_{6j}$ | Secret $l_{6j}$ | Public $m_{6j}$ | Public $n_{6j}$ | Public $v_{6j}$ |
|---|---|---|---|---|---|
| User – $S_1$ | 2 | 7 | 25 | 9 | 34 |
| User – $S_2$ | 5 | 19 | 18 | 8 | 33 |
| User – $S_3$ | 7 | 4 | 34 | 25 | 32 |
| User – $S_4$ | 11 | 7 | 25 | 4 | 16 |
| User – $S_5$ | 13 | 19 | 18 | 36 | 45 |
| User – $S_7$ | 12 | 2 | 9 | 12 | 17 |

(i). User- $S_7$ makes the following table.

| VALUE \ USER | Secret $h_{7j}$ | Secret $l_{7j}$ | Public $m_{7j}$ | Public $n_{7j}$ | Public $v_{7j}$ |
|---|---|---|---|---|---|
| User – $S_1$ | 11 | 19 | 18 | 4 | 5 |
| User – $S_2$ | 13 | 7 | 25 | 36 | 27 |
| User – $S_3$ | 15 | 15 | 42 | 42 | 26 |
| User – $S_4$ | 17 | 16 | 32 | 2 | 3 |
| User – $S_5$ | 14 | 8 | 28 | 16 | 14 |
| User – $S_6$ | 12 | 16 | 32 | 12 | 2 |

*Partial Signature Generation by any t Users*

If any four members $S_2, S_4, S_6$, and $S_7$ are agree to sign a message $m$ for a person $R$, possessing $x_R = 7$, $y_R = 25$, then the signature generation has the following steps.

(a). $S_2$ randomly selects $K_{2_1} = 11, K_{2_2} = 13$ and computes $u_2 = 1, v_2 = 4$ and $w_2 = 17$.

(b). $S_4$ randomly selects $K_{4_1} = 10, K_{4_2} = 12$ and computes $u_4 = 4, v_4 = 17$ and $w_4 = 9$.

(c). $S_6$ randomly selects $K_{6_1} = 14, K_{6_2} = 17$ and computes $u_6 = 24, v_6 = 14$ and $w_6 = 6$.

(d). $S_7$ randomly selects $K_{7_1} = 18, K_{7_2} = 5$ and computes $u_7 = 6, v_7 = 6$ and $w_7 = 25$.

(e). Each user computes $U_S = 16, R_R = 25, W_S = 14$ and $R_S = h(25, m) = 7$. (let).

(f). $S_2$ recovers his shares $l_{12} = 4, l_{32} = 5, l_{52} = 22$ and $C_2 = 1, MS_2 = 8, s_2 = 15$.

(g). $S_4$ recovers his shares $l_{14} = 5, l_{34} = 1, l_{54} = 5$ and $C_4 = 11, MS_4 = 6, s_4 = 10$.

(h). $S_6$ recovers his shares $l_{16} = 15, l_{36} = 4, l_{56} = 21$ and $C_6 = 9, MS_6 = 15, s_6 = 15$.

(i). $S_7$ recovers his shares $l_{17} = 16, l_{37} = 17, l_{57} = 11$ and $C_7 = 3, MS_7 = 17, s_7 = 8$.

*Partial Signature Verification and Signature Generation by DC*

(a) *DC* verifies each partial signature. For example for user $S_2$, $s_2 = 15, v_2 = 4, m_{12} = 34, m_{32} = 8, m_{52} = 16, y_{S_2} = 37, R_S = 7, C_2 = 12$, and check $3^{16} \stackrel{?}{\equiv} 4 \cdot 37^7 (34 \cdot 8 \cdot 16)^7 \mod 47$. This holds. Similarly, he checks other partial signatures.

(b). *DC* computes a group value $S_S = 2$ and sends $\{2, 16, 14, m\}$ as signature of the group $S$ for the message $m$.

*Signature Verification by the Person R*

(a). The verifier $R$ computes a verification value $E = 12$.

(b). The verifier $R$ can recovers the values $R_R = 25$ and $R_S = 7$.

(c). The verifier $R$ uses the congruence $3^2 \stackrel{?}{\equiv} 25 \cdot (12.25)^7 \mod 47$ to check the validity of the signature. This congruence holds, so the group signature $\{2, 16, 14, m\}$ is valid signature of the group $S$ on the message $m$ for the person $R$.

*Proof of Validity by R to any third Party C*

(a). $R$ computes $\mu = 32, R_R = 25$ and sends $\{25, 12, 2, 16, m, 32\}$ to C.

(b). C recovers $R_S = 7$ and check the concurrence $3^2 \stackrel{?}{\equiv} 25 \cdot (12.25)^7 \mod 47$ for the validity of the signature. This holds; so, C goes to the next steps.

(c). $R$ in a zero knowledge fashion proves to C that $\log_{16} 32 = \log_3 25$ as follows:-

- C chooses random $u = 13, v = 15$ and computes $w = 9$ and sends $w$ to $R$.

- $R$ chooses random $\alpha = 11$ computes $\beta = 36$, $\gamma = 16$ and sends $\beta$, $\gamma$ to C.
- C sends $u, v$ to $R$, by which $R$ can verify that $w = 9$.
- $R$ sends $\alpha$ to C, by which she can verify that $\beta = 36$ and $\gamma = 16$.

## 5. Remarks

In this paper, we have proposed a **Directed –Threshold Multi - signature Scheme without SDC.** In this scheme,

- **Each shareholder works as a SDC** to generate his secret key and distribute the corresponding secret shares to other shareholders.
- There is a **designated combiner** *DC* who takes the responsibility to collect and verify each partial signature and then produce a group signature, **but no secret information is associated with the *DC*.**
- Any malicious set of signers cannot impersonate any other set of signers to forge the signatures. In case of forgery, it is possible to trace the signing set.
- Any *t* or more shareholders acting in collusion cannot conspire to reconstruct the group secret key by providing their own secret shares.

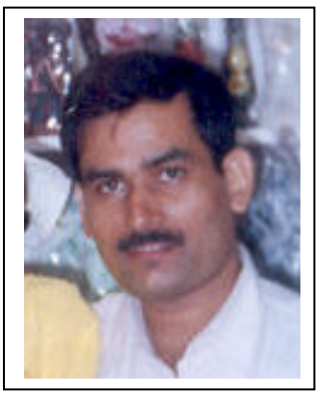

**Manoj Kumar** received the B.Sc. degree in mathematics from Meerut University Meerut, in 1993; the M. Sc. in Mathematics (Goldmedalist) from C.C.S.University Meerut, in 1995; the M.Phil. (Goldmedalist) in *Cryptography*, from Dr. B.R.A. University Agra, in 1996, completed the Ph.D. in applied Mathematics, in 2003. He also taught applied Mathematics at DAV College, Muzaffarnagar, India from Sep 1999 to March 2001; at S.D. College of Engineering & Technology, Muzaffarnagar, and U.P., India from March 2001 to Nov 2001; at Hindustan College of Science & Technology, Farah, Mathura, continue since Nov 2001. He also qualified the *National Eligibility Test* (NET), conducted by *Council of Scientific and Industrial Research* (CSIR), New Delhi- India, in 2000. He is a member of Indian Mathematical Society, Indian Society of Mathematics and Mathematical Science, Ramanujan Mathematical society, and Cryptography Research Society of India. His current research interests include Cryptography, Numerical analysis, Pure and Applied Mathematics.